\documentclass[12pt]{article}
\usepackage{latexsym}
\usepackage{amsmath}
\usepackage{wasysym}
\usepackage{amsfonts}

 \csname
@addtoreset\endcsname{equation}{section}

\def\pe{\prime}
\def\3s{{s \choose 3}}
\def\4s{{s \choose 4}}
\def\5s{{s \choose 5}}
\def\6s{{s \choose 6}}

\def\12{\frac{1}{2}}
\def\fr{\frac}
\def\ft{\footnote}
\let\la=\label

\def\pr{\partial}

\def\prd{\partial \cdot}

\def\ra{\rightarrow}

\def\be{\begin{equation}}
\def\ee{\end{equation}}
\def\bea{\begin{eqnarray}}
\def\eea{\end{eqnarray}}
\def\ba{\begin{array}}
\def\ea{\end{array}}
\def\bec{\begin{center}}
\def\ec{\end{center}}

\def\scs#1{\section{\sc #1}}

\def\a{\alpha}

\def\d{\delta} 

\def\e{\epsilon}

\def\th{\theta}

\def\l{\lambda}
\def\L{\Lambda}
\def\m{\mu}
\def\n{\nu}

\def\r{\rho}

\def\vf{\varphi}
\def\c{\chi}

\def\eb{{\bar\epsilon}}

\def\lb{\bar{\lambda}}
\def\cb{\bar{\chi}}
\def\pb{{\bar\psi}}

\def\dsll{\not {\! \pr}}
\def\psisl{\not {\! \! \psi}}

\def\ssl{\not {\! \cal S}}

\def\ssl{\not {\! \cal S}}

\def\cF{{\cal F}}

\def\cJ{{\cal J}}

\def\cL{{\cal L}}

\def\cR{{\cal R}}
\def\cS{{\cal S}}

\begin{document}
\begin{flushright}
%hep-th/yymmddd \vskip 8pt {\today}
\end{flushright}

\vspace{20pt}

\begin{center}

%%%%%%%%%%%%%%%%%%%%%%%%%%%%%%%%%%%%%%%%%%%%%%%%%%%%%%%%%%%%%%%%%%%%

{\Large\sc String theory triplets and higher-spin curvatures}\\

%%%%%%%%%%%%%%%%%%%%%%%%%%%%%%%%%%%%%%%%%%%%%%%%%%%%%%%%%%%%%%%%%%%%

\vspace{30pt} {\sc Dario Francia} \\ \vspace{5pt} 
AstroParticule et Cosmologie (APC) \footnote{Unit\'e mixte de Recherche du CNRS (UMR 7164.)}\\
Universit\'e Paris VII - Campus Paris Rive Gauche \\
10, rue Alice Domon et L\'eonie Duquet \\ F-75205 Paris Cedex 13\
FRANCE
\\ e-mail:
{\small \it francia@apc.univ-paris7.fr}
\vspace{10pt}

%%%%%%%%%%%%%%%%%%%%%%%%%%%%%%%%%%%%%%%%%%%%%%%%%%%%%%%%%%%%%%%%%%%%

\vspace{30pt} {\sc\large Abstract}\end{center}
Unconstrained local Lagrangians for higher-spin gauge theories are bound to involve auxiliary fields, whose integration in the partition function generates geometric, effective actions  expressed in terms of curvatures. When applied to the triplets, emerging from the tensionless limit of open string field theory, the same procedure yields interesting alternative forms of geometric Lagrangians, expressible for both bosons and fermions as squares of field-strengths. This shows that higher-spin curvatures might play a role in the dynamics, regardless of whether the Fronsdal-Labastida constraints are assumed or forgone.

\vfill
\setcounter{page}{1}

%\pagebreak

%\tableofcontents

\newpage

%%%%%%%%%%%%%%%%%%%%%%%%%%%%%%%%%%%%%%%%%%%%%%%%%%%%%%%%%%%%%%%%%%%%%

\scs{Introduction and Summary}\label{sec1}

%%%%%%%%%%%%%%%%%%%%%%%%%%%%%%%%%%%%%%%%%%%%%%%%%%%%%%%%%%%%%%%%%%%%%

 The investigation of the properties of string theory, in a properly defined low-tension limit, should provide crucial information on its massless phase with higher-spin gauge symmetries, long conjectured to describe its high energy regime \cite{ouvry, gross, fradkin}. However, while several options exist for the definition of this limit, the interpretation of the resulting interacting theory  is still not fully clarified\ft{For instance, one may choose to consider the limit  either in the classical action \cite{karlind}, or in  the scattering amplitudes, at the full quantum level \cite{amatigross}. Consistency is dubious in flat space-time, since interacting massless higher-spins  are usually supposed to require a non-vanishing cosmological  constant \cite{fradkinvas}. For a recent alternative approach to massless higher-spins in \emph{tensile} strings see \cite{polyakov}.}  \cite{tensionless1, tensionless2}.

 The situation simplifies if one confines the attention to the free sector of the Lagrangian of 
open bosonic string field theory \cite{sftheory}. In this framework in fact, as first suggested  in \cite{ouvry, triplets}, 
properly rescaled Virasoro generators satisfy a simplified algebra, in the  limit $\a^{\, \pe} \ra \infty$,
without  central charge. The corresponding, identically nilpotent, BRST operator allows to define
in any space-time dimension gauge-invariant Lagrangians for massless, but otherwise arbitrary representations of the Poincar\'e group \cite{fs2, st} (see also  \cite{bekbuch, ftrev} for reviews). 
In particular, for the leading Regge trajectory associated to symmetric fields, the 
equations of motion and the  gauge transformations for the corresponding systems, 
collectively referred to as  \emph{triplets}, read
\begin{align} \label{triplet0}
&\Box \, \vf \, = \, \pr \, C \, , & &&   &\d \, \vf \, = \, \pr \, \L \, , & \nonumber \\
&C \, = \, \prd \, \vf \, - \, \pr \, D \, ,& && &\d \, C \, = \, \Box \, \L \, ,& \\
&\Box \, D \, = \, \prd C \, ,  & &&  &\d \, D \, = \, \prd \, \L \, . \nonumber &
\end{align}
where $\vf$, $C$ and $D$ have  ranks $s$, $s-1$ and $s-2$ respectively, in a notation where
symmetrised indices are not displayed\ft{We use the ``mostly-plus'' space-time metric in $d$ dimensions.  
Unless otherwise stated, in all expressions symmetrised indices are implicit, and in the product of different 
tensors symmetrization is understood with no weight factors. Lorentz traces are denoted by ``primes'', 
while the symbol ``$\prd$'' stands for a divergence. A full list of identities can be found in 
\cite{fs1, fs2}. To the purposes of this paper the only needed one is
$
 \partial^{\, p}  \partial^{\, q}  \,  = \, {p+q \choose p} 
\partial^{\, p+q} \, 
$.}.
Fermionic triplets can be derived similarly, starting from the rescaling of the Virasoro generators in  the open superstring \cite{fs2, st}.

The triplet equations \eqref{triplet0} provide a description of symmetric gauge fields of any spin \cite{hsp} somehow alternative to Fronsdal's one \cite{fronsdal}. The main difference with respect to \cite{fronsdal} is that neither fields nor parameters are subject to trace-constraints, while in Fronsdal's bosonic case it is assumed
that the gauge parameter $\L$  be traceless and the gauge field $\vf$ be doubly-traceless. 
In addition,  eqs.  \eqref{triplet0} can be shown to propagate 
irreducible modes of spin $s$, $s-2$, $s-4$, $\cdots$, down to  $1$ or $0$, according to whether $s$ is odd or even
\cite{fs2, st}, to be contrasted with the wave equation of Fronsdal, that propagates irreducible spin $s$.  Similar considerations apply
to mixed-symmetry triplets \cite{st}, to be compared with the 
corresponding formulation for irreducible fields of arbitrary symmetry given by Labastida
\cite{labastida}, that involves proper generalisations of the Fronsdal constraints.
Absence of  trace constraints and propagation of sets of irreducible representations are related issues, and indeed
in \cite{ftcurrents} it was shown that  eqs. \eqref{triplet0}, after 
elimination of the field $C$, can be recast as a set of decoupled Fronsdal equations for the corresponding 
representations. 

 At any rate,  the absence of (manifest) constraints in string-inspired descriptions of higher-spin 
gauge fields stimulated  the investigation of theories
possessing a wider gauge symmetry than the one assumed in the constrained setting. 
An additional significant motive was related to the fact that the Fronsdal-Labastida constraints appear in principle
at odds with the structural  properties of higher-spin curvatures.

 Higher-spin generalisations of the Maxwell and the Riemann linearised field-strengths were 
introduced in \cite{weinberg65, dwf} for symmetric (spinor-)tensors, and were first investigated for mixed-symmetry
fields in \cite{dvh}. They represent the best candidates to describe properties of higher-spin geometry 
in metric form, at the linearised level. On the other hand, since they contain a number of derivatives equal to 
the spin, it was long assumed that they could play no dynamical role for spin $s \geq 3$,  
in agreement with the fact that their structure bears no apparent relation with the Fronsdal-Labastida constraints, 
and thus with the usual construction of higher-spin Lagrangians\ft{In particular, curvature tensors  are  gauge-invariant under unconstrained transformations of the potential.}.

 The problem was then reconsidered in  \cite{fs1, fs2, bbl, dmh, fms1, dario07}, where it was shown
how to define Lagrangians and equations of motion built out of curvatures, describing 
the free propagation of \emph{irreducible}, massless fields of any spin and symmetry.
Those results generalised the works of Fronsdal and Labastida, 
that were shown to describe partially gauge-fixed forms of the corresponding geometrical theories.
On the other hand, the resulting picture  turned out to be rather unconventional, 
involving either non-local Lagrangians, or high-derivative, non Lagrangian, equations of motion. 
Thus, in order to try and reproduce the features of those theories in a more standard
setting,  the related problem of formulating \emph{unconstrained, local} Lagrangians 
for irreducible higher spins was also investigated.

 The simplest solution to this problem\ft{Unconstrained formulations of irreducible higher-spins
had been previously investigated in \cite{bpt}.}  was given in a series of works \cite{fs3, fms1, dario07, fms2, cfms1, cfms2, andrearev},
providing minimal local Lagrangians for irreducible gauge fields of any spin and symmetry.
In particular, for the case of symmetric bosons, the basic result of both local and non-local formulations has been to show that the Lagrangian equations of motion
can be always reduced to the ``compensator form''
\be \label{compensator}
\cF \, (\vf)\, = \, 3 \, \pr^{\, 3} \, \a \, ,
\ee
where $\cF\,  (\vf)$ is the kinetic tensor of the Fronsdal theory, while the compensator $\a$ is either an independent field, in local formulations, 
or a non-local tensor function of $\vf$, in the geometric setting.
In both cases it transforms proportionally to the trace of the gauge parameter, so that, once 
\eqref{compensator} is reached, the conventional Fronsdal equation, $\cF \,  (\vf)\, = 0$, can be immediately
recovered\ft{The compensator equation \eqref{compensator} can be recovered  
from a consistent truncation of  \eqref{triplet0}, imposing the additional condition
$\vf^{\, \pe} - 2\, D = \pr \, \a$ (where $\vf^{\, \pe}$ denotes the trace of $\vf$), 
which is tantamount to declaring that all lower-spin modes in $\vf$ are pure gauge \cite{fs2, st}.
The reduction of the triplet Lagrangian to irreducible spin $s$ is discussed in \cite{triplobuch}.}. 
On the other hand, despite the unconstrained gauge invariance underlying the triplets
of \cite{fs2, st}, to date there has been no attempt to uncover possible relations of the 
corresponding Lagrangians with the curvatures of \cite{weinberg65, dwf}.

 In this letter we would thus like to investigate the geometrical meaning of the \emph{reducible} description
of gauge fields provided by the triplets, limiting ourselves to the case of symmetric 
(spinor-)tensors in metric form\ft{A detailed discussion of triplets in generalised vielbein formalism can be found in \cite{sorokvas}.}. 
The underlying idea is simple: the triplet Lagrangian is identically
gauge invariant under unconstrained gauge transformations, in the same fashion as the higher-spin curvatures are.
It involves auxiliary fields, but once the gaussian integration over those fields
is explicitly performed  in the partition function,  the resulting ``effective'' Lagrangian would involve 
the field $\vf$ alone, and  consequently  it must be possible to express it  in terms 
of curvatures. The same  idea was already exploited in \cite{dariocrete}, in order to clarify
the relation between local and geometric theories for the irreducible case.

 The main outcome of this investigation are the Lagrangians \eqref{Maxwellbose} and \eqref{Maxwellfermi}, 
that represent the geometric counterparts of  bosonic and  fermionic triplet Lagrangians of
\cite{fs2, st}, respectively. The resulting equations of motion take a remarkably simple form when expressed in terms of curvatures: in the bosonic case, here displaying indices for clarity, they are
\be \label{eomboseintro}
\fr{1}{\Box^{s - 1}} \, \pr^{\, \m_1} \, \cdots \, \pr^{\, \m_s} \, \cR_{\, \m_1 \cdots \m_s, \, \n_1 \cdots  \n_s}  \, 
(\vf) \, = \, 0 \, ,
\ee
where $\cR_{\, \m_1 \cdots \m_s, \, \n_1 \cdots  \n_s} \, (\vf)$ is the curvature tensor of \cite{dwf}, whose definition
in terms of  the symmetric gauge potential $\vf_{\, \m_1 \cdots \m_s}$ is recalled in Section \ref{sec2.1}.
The corresponding equations for fermionic, rank-$s$, spinor-tensors $\psi_{\, \m_1 \cdots \m_s}$ are
\be \label{eombosefermiintro}
\fr{\dsll}{\Box^{s}} \,\pr^{\, \m_1} \, \cdots \, \pr^{\, \m_s} \, \cR_{\, \m_1 \cdots \m_s, \, \n_1 \cdots  \n_s}  \, 
(\psi) \, = \, 0 \, ,
\ee 
and thus can be formally obtained from the bosonic ones, interpreting the gauge potential as carrying an additional spinor index, 
while formally acting on \eqref{eomboseintro} with the operator $\fr{\not \, \pr}{\Box}$. We would like to stress that no such a simple relation is found, in the irreducible case, between bosonic and fermionic equations, either constrained or unconstrained. To summarise, our results suggest the following remarks and speculations:
\begin{itemize}
 \item basically, what we show in this work is that curvatures can be used to describe the propagation of 
\emph{several} Fronsdal fields, properly combined so as to reconstruct full gauge invariance of an unconstrained gauge potential. In this sense, we are led to conclude that presence or absence of constraints is probably not the clue to distinguish whether higher-spin curvatures may or may not play a role.
\item The triplets coming from the tensionless limit of string field theory provide a relevant example to this effect
(not necessarily the only one): the integration over the auxiliary fields \emph{defines}  geometric Lagrangians and equations of motion which are alternative with respect to those investigated up to now. We observe that in  \eqref{eomboseintro} and \eqref{eombosefermiintro} \emph{traces} of the curvatures are absent, while it is known \cite{damdes, dvh, bbl} that the condition that the trace of the curvature be zero is equivalent to impose a compensator-like equation of the form \eqref{compensator},  implying propagation of \emph{irreducible} spin $s$\ft{Analogously, the form of the non-local equation \eqref{eombose}
suggests that its local, higher-derivative, non-Lagrangian counterpart should be obtained by the condition
of vanishing divergence of the curvature. }. 
\item As a reflection of the simplicity of the triplet Lagrangians \eqref{triplet} and \eqref{tripletfermi},  their geometric counterparts \eqref{Maxwellbose} and \eqref{Maxwellfermi} are much simpler than the corresponding ones for irreducible fields, presented in \cite{fs1, fs2, fms1, dariocrete}. It is tempting to speculate that this simplification could play a role at the interacting level, although we are not in the position to add much in this respect, if not for a few 
conjectures  proposed in the following remark. 
 
 \item In all their known realizations, interacting higher-spin theories involve auxiliary fields. In view of 
what we discuss here and in \cite{dariocrete}, we judge it  conceivable that interacting Lagrangians for gauge fields of any spin might admit an interpretation in terms of non-linearly deformed curvatures, once the auxiliary fields (or maybe a proper subset of them) are integrated away. Reversing the logic, one could  look first for consistent deformations of non-local theories formulated in terms of curvatures, possibly suggested by some geometrical intuition. If any such programme is at all viable, we expect the extreme simplicity of the Lagrangians
here presented to be of some help. The resulting theory  could then be converted in more conventional terms, trading non-localities for auxiliary fields, with the bonus of additional insight into the meaning of the interactions themselves.
\end{itemize} 

  We investigate the geometric form of the bosonic triplets in Section \ref{sec2}, while  the fermionic one 
is discussed  in Section \ref{sec3}.
In both cases we first rederive the triplet Lagrangians in a simple way, alternative to the BRST construction 
of \cite{fs2, st, ftrev}. In Section \ref{sec4} we comment on the equations of motion and on the corresponding current exchanges. The details of the inversion of the operator in the gaussian
integration over the auxiliary fields are discussed in the Appendix.

%%%%%%%%%%%%%%%%%%%%%%%%%%%%%%%%%%%%%%%%%%%%%%%%%%%%%%%%%%%%%%%%%%%%%

\section{Bosonic triplets}\label{sec2}

%%%%%%%%%%%%%%%%%%%%%%%%%%%%%%%%%%%%%%%%%%%%%%%%%%%%%%%%%%%%%%%%%%%%%

 The triplet Lagrangian was first written in \cite{fs2, st}, following the procedure sketched in the Introduction.
 A simple alternative derivation obtains starting with the trial Lagrangian
\be \label{L0}
\cL_0 \, = \, \12 \, \vf \, \Box \, \vf \, ,
\ee
and compensating its variation under $\d \, \vf = \pr \, \L$ by as many
auxiliary fields as independent functions of the parameter appear in intermediate steps.
In general, in fact, the requirement that the full Lagrangian be gauge invariant does not fix
uniquely the form of the additional terms to be introduced, and the resulting theory 
will possess different properties according to whether or not one tries  to keep to a minimum the number 
of \emph{off-shell} components\ft{Keeping to a miminum the number of off-shell components uniquely leads to
the Fronsdal theory. Indeed, assuming that the kinetic tensor contain $\Box \, \vf$, and 
compensating its variation preserving locality, but without introducing any auxiliary fields, leads to 
the Fronsdal tensor
\be \la{fron}
\cF = \Box \vf \, - \, \pr \, \prd \vf \ + \, \pr^{\, 2} \, \vf^{\, \pe}\, ,
\ee with constrained gauge invariance: $\L^{\, \pe} \, \equiv \, 0$. The double trace constraint is then needed
for the construction of a gauge-invariant Lagrangian.}.

 In the case of the triplets, in order to compensate the variation of \eqref{L0}, 
we first introduce a new field $C$ transforming as $\d \,C = \Box \,\L$ and write
$
\cL_1 \, = \, \cL_0 \, + \, s \, \prd \, \vf \, C \, .
$
In the variation of $\cL_1 $, on the other hand, two terms appear:
\be
\d \, \cL_1 \, = \, s \, \Box \, \L \, C  \, - \, s \, {s \choose 2} \, \prd \L \, \prd C \, ,
\ee 
the first of which can be compensated by a quadratic term in $C$, while for the second we need (according to the spirit of this derivation) a new field $D$ s.t. $\d \, D = \prd \L$. In this way, we finally recover the triplet Lagrangian \cite{fs2, st}
\be \label{triplet}
\begin{split}
\cL  = &  \, \12 \,  \vf \,  \Box \, \vf
 - \, \12 \, s \, C^{\, 2} \, - \,  {s \choose 2}   \, D \,  \Box \, D \,
 + \, s\, \prd \vf \, C \, +\, 2 \, {s \choose 2} \,D \, \prd  C  \, . 
\end{split}
\ee
As already stressed, differently from the
local Lagrangians of \cite{bpt, fs3, fms1, triplobuch, dario07, cfms1, cfms2}, in this case unconstrained gauge invariance
is related to \emph{reducible} higher-spin propagation. The task of decomposing \eqref{triplet} 
in a sum of constrained, Fronsdal Lagrangians, each for any of the irreps propagating in \eqref{triplet0}, 
was undertaken in \cite{ftcurrents} \ft{More precisely, the decomposition is performed
for the Lagrangian resulting from \eqref{triplet}, after $C$ is removed using its equation of motion.
In \eqref{triplet}, taken at face value, there are still \emph{more} off-shell components than 
in the corresponding sum of Fronsdal Lagrangians.}. Here we would like to exploit its unconstrained 
gauge invariance to establish a connection with the higher-spin curvatures of \cite{dwf}.

%%%
\subsection{Effective Lagrangians and higher-spin curvatures} \label{sec2.1}
%%%
Similarly to what we  did in \cite{dariocrete} for the irreducible case, we consider the gaussian partition function for the theory defined by the Lagrangian \eqref{triplet}, and integrate over the fields $C$ and  $D$  (or, equivalently, we can solve for $C$ and $D$ from their equations of motion and then substitute in $\cL$)  obtaining in this way the effective Lagrangian for $\vf$
\be \label{Zcd}
\begin{split}
\cL_{\, eff}\, (\vf) \, = \,  \12  \,  \vf \, (\Box \, - \, \pr \, \pr \, \cdot) \, \vf \, + \, \12 \, {s \choose 2} \, 
\prd \prd \vf \,  (\Box \, + \, \12 \, \pr \, \pr\, \cdot)^{\, -1} \, \prd \prd \vf  \, ,
\end{split}
\ee
that we would like to make more explicit, computing the 
inverse of the operator
\be \label{A}
A  =  \, (\Box \, + \, \12 \,  \pr \, \pr\, \cdot) \, .
\ee
The details of the calculations are presented in the Appendix. Here we just observe that
the result depends on the rank of the tensors on which  $ A^{\, -1} $ is supposed to act.
Thus for instance on $T_{\, (1)}$, the space of rank-$1$ tensors, one obtains
\be
A ^{\, -1}_{\, (1)} \, = \,  \, \Box^{\, - 1} \, (1 \, - \, \fr{1}{3} \, \fr{\pr}{\Box} \, \pr\, \cdot) \, , 
\ee
and the resulting effective Lagrangian for the spin-$3$ case reads
\be
\cL_{eff}\, (\vf) \, = \, \12 \, \{ \vf \, \Box \, \vf \, + \, 3 \, \prd \vf \, \prd \vf \, + \, 
3 \, \prd \prd \vf \, \fr{1}{\Box} \,  \prd \prd \vf  \, + \,  \prd \prd \prd \vf \, \fr{1}{\Box^2} \,  \prd \prd \prd \vf \}\, .
\ee
More generally, on $T_{\, (k)}$, the space  of tensors of rank $k$, the inverse of $A$ is given by 
\be \label{inverseA}
A^{\, -1}_{\, (k)} \, = \, \fr{1}{\Box} \, \{1\, + \, \sum_{m = 1}^k \, (-1)^m \, 
\fr{m!}{2^{\, m} \, \prod_{l = 1}^m \, (1\, + \, \fr{l}{2})} \, \fr{\pr^{\, m}}{\Box^{\, m}} \, \pr\, \cdot^m\}\, ,
\ee
so that, after some manipulations,  the effective Lagrangian for the general case of spin $s$ can be written in 
one of the two equivalent forms
\be \label{effectiveL}
\begin{split}
\cL_{eff}\, (\vf) \, & = \, \12 \, \sum_{m = 0}^{s} \, {s \choose m}\, 
\prd^{\, m} \vf \fr{1}{\Box^{\, m - 1 }} \, \prd^m \vf \\
&= \, \12 \, \vf \, \{\Box \, \vf \, + \,  \sum_{m = 1}^{s} \, (-1)^m \, 
\fr{\pr^{\, m}}{\Box^{\, m - 1}} \, \prd^m  \vf \} \, ,
\end{split}
\ee
where in particular the term in curly brackets is identically divergenceless and gauge invariant (thus granting for gauge invariance of 
$\cL_{eff}\, (\vf)$), and indeed can be directly related to the curvatures of \cite{dwf}, as we now discuss. 

 To establish a  connection with higher-spin geometry we recall the definition of curvatures given in \cite{dwf} for a symmetric tensor
of rank $s$:  
\be \label{curvatures}
\begin{split}
&\cR^{\, (s)}_{\, \m_s, \,  \n_s} \, 
= \, \sum_{k=0}^{s}\fr{(-1)^{k}}{
\left(
{{s} \atop {k}}
\right)
}\ 
\pr^{s-k}_{\, \m}\pr^{k}_{\, \n} \vf_{\, \m_k,\,\n_{s-k}} \, ,
\end{split}
\ee
in a notation (see \cite{dario07}, Sec. $2.1.1$) where indices denoted by the same letter are to be understood as being completely symmetrised, 
without normalization factors, with the minimum number of terms required, while no symmetry between different
groups is assumed. Moreover,
$ 
\pr^{\, k}_{\, \r} 
$
denotes the product of $k$ uncontracted gradients. 

Using \eqref{curvatures} it is possible to express  the effective Lagrangian 
\eqref{effectiveL} in terms of curvatures, and the result takes the rather compact 
form\ft{For the generic term in \eqref{curvatures} the computation gives
$
\pr^{\m_1}\cdots  \pr^{\m_s} \{\pr^{s-k}_{ \m}\pr^{k}_{ \n} \vf_{ \m_k,\,\n_{s-k}}\}  =  
{s \choose k}  \pr^{\, k} \Box^{s - k}  \prd^k \vf \, ,
$
where in the r.h.s. the symmetrised indices belonging to the $\n$-group are omitted.   
In particular, the number of contributions to the final result is simply given by
the number of ways in which one can saturate the $s - k$ gradients in the $\m$ indices with the 
corresponding $s$ divergences.}
\be \label{geomL}
\cL_{eff}\, (\vf) \, = \, \12 \, \vf \, \fr{1}{\Box^{s - 1}} \, \prd^{\, s} \, \cR^{\, (s)} \, ,
\ee 
where the $s$ divergences are computed with respect to indices belonging to the first group\ft{Gauge
invariance of  \eqref{geomL}, manifest in the form \eqref{Maxwellbose},
can be also understood in terms of the group theoretical properties of the curvatures
\eqref{curvatures}. Indeed, they define irreducible, two-rows $GL(D, \mathbb{R})$ tensors, so that symmetrization 
of one index in the $\n$-group with the indices of the $\m$ group gives zero. For this reason the $(s + 1)$-th divergence
of $\cR^{\, (s)}$ vanishes identically, if $s$ divergences are computed with respect to the same group.}, 
so as  to avoid the appearance of a  spin-dependent sign related to the exchange property $\cR^{\, (s)}_{\, \m_s, \,  \n_s}
= (-1)^{\, s} \, \cR^{\, (s)}_{\, \n_s, \,  \m_s}$. Alternatively, it is possible to integrate by parts
the divergences thus obtaining for the Lagrangians  \eqref{geomL} the suggestive Maxwell-like forms\ft{I am especially grateful
to X. Bekaert and E. Joung for drawing to my attention this rewriting of  the Lagrangians  \eqref{geomL}.}
\be \label{Maxwellbose}
\cL_{eff}\, (\vf) \, =  \, \fr{(-1)^{\, s}}{2\, (s + 1)} \, 
\cR^{\, (s)}_{\, \m_s, \,  \n_s} \, \fr{1}{\Box^{s - 1}} \, \cR^{\, (s)\,  \m_s, \,  \n_s} \, .
\ee 
 Formula \eqref{Maxwellbose}, together with its fermionic counterpart  \eqref{Maxwellfermi} to be derived
in Section \ref{sec3},  represent the main result of this paper\ft{For the spin-$2$ case, infrared modifications of the Einstein-Hilbert
action involving inverse powers of the D'Alembertian operator have been proposed in \cite{barv}. 
In the MacDowell-Mansouri-Stelle-West  formulation of gravity \cite{mm, sw}  the Lagrangian is written as a square of curvatures
(in the sense of the wedge product), in the Cartan-Weyl formalism. On the other hand, the resemblance of that result and of its generalisations to free higher-spins, as reviewed by Bekaert, Cnockaert et al. in [17],  with the Lagrangian \eqref{Maxwellbose} is not so direct. Indeed, once the MacDowell-Mansouri-like system is reduced to metric-form the resulting Lagrangian reproduces the Fronsdal one, describing irreducible spin $s$ in terms of constrained fields, and for this reason it cannot be equivalent to \eqref{Maxwellbose}.}.

%%%%%%%%%%%%%%%%%%%%%%%%%%%%%%%%%%%%%%%%%%%%%%%%%%%%%%%%%%%%%%%%%%%%%

\section{Fermionic triplets}\label{sec3}

%%%%%%%%%%%%%%%%%%%%%%%%%%%%%%%%%%%%%%%%%%%%%%%%%%%%%%%%%%%%%%%%%%%%%

 Fermionic triplets were also discussed in \cite{fs2, st}, and in particular were linked to the tensionless 
limit of the open superstring in \cite{st}. An alternative derivation of their Lagrangian  obtains mimicking
what we did for bosons in Section \ref{sec2}. One starts with the basic quadratic form
\be
\cL_0 \, = \, i \, \pb \, \dsll \, \psi \, ,
\ee 
where $\psi$ is a rank-$s$ spinor-tensor, whose first-order variation under $\d \psi = \pr \, \e$, $\d \pb = \pr \, \eb$ 
can be compensated introducing an auxiliary field $\c$ transforming as $\d \, \c \, = \, {\not \! \pr} \,\e$,
leading to the new trial Lagrangian
\be \label{L1}
\cL_1 \, = \,  \cL_0 \, - \, i \, (\pr \, \cb \, \psi \, + \, \pb \, \pr \, \c) \, .
\ee
Again, following the idea of this derivation, we balance the divergences of the parameter appearing in the
variation of \eqref{L1} introducing a new field $\l$, with
$\d \, \l \, = \, \prd \e$, finally obtaining the Lagrangian of the fermionic triplet \cite{fs2, st}:
\be \label{tripletfermi}
\begin{split}
\cL &=  \, i \, \bar{\psi} \, \dsll \,\psi \,  - \, i \, s \, \bar{\chi} \, \dsll \, \chi 
 \, - \, 2 \, i \, {s \choose 2} \, \bar{\l} \, \dsll \, \l \, 
 + \, \{i\, s \, \cb \, \prd \psi  \, + \, 2 \, i\, {s \choose 2} \, \lb \, \prd \c  +  h. c.\} \, .
\end{split}
\ee

%%%
\subsection{Effective Lagrangians and higher-spin curvatures}
%%%

 Performing the  fermionic gaussian integration over $\cb$ and $\c$ and then over  $\lb$ and $\l$ we obtain 
the effective Lagrangian for the fermionic spin-$(s + \12)$ triplet
\be \la{lagreffferm}
\cL_{eff}\,  (\psi, \pb) \, = \, i \, \bar{\psi} \, \dsll \,\psi \, + \, i \, s \, \prd \pb \, \fr{\dsll}{\Box} \, \prd \psi \,+
\, i \,  \prd \prd \pb \, [\dsll \, (\Box \, + \, \12 \, \pr \, \prd)]^{\, -1} \, \prd \prd \psi \, ,
\ee
where the strong resemblance with the bosonic result, in particular as given in  \eqref{Zcd}, is already manifest. 
Corresponding similarities  are met when expressing  $\cL_{eff}\,  (\psi, \pb)$ in terms of curvatures. 
Indeed, making use of \eqref{inverseA}, we can write  \eqref{lagreffferm} in the explicit forms:
\be \label{effectiveFerm}
\begin{split}
\cL_{eff}\, (\psi, \pb)  \, & =  \, i \, \sum_{m = 0}^{s} \, {s \choose m}\, 
\prd^{\, m} \pb \fr{\dsll}{\Box^{\, m}} \, \prd^m \psi \\
& = \, i \, \pb \, \dsll\, \{\psi \, + \,  \sum_{m = 1}^{s} \, (-1)^m \, 
\fr{\pr^{\, m}}{\Box^{\, m}} \, \prd^m \psi \} \, , 
\end{split}
\ee
where again, as for the case of bosons, the second can be directly expressed in terms of the curvatures \eqref{curvatures} as
\be \label{geomLferm}
\cL_{eff}\, (\psi, \pb) \, = \, i \, \pb \, \fr{\dsll}{\Box^{s }} \, \prd^{\, s} \, \cR^{\, (s)} \, ,
\ee 
or, equivalently, in the more inspiring form
\be \label{Maxwellfermi}
\cL_{eff}\,(\psi, \pb) \, =  \, \fr{(-1)^{\, s} }{s + 1} \, i\, 
\bar{\cR}^{\, (s)}_{\, \m_s, \,  \n_s} \, \fr{\dsll}{\Box^{s}} \, \cR^{\, (s)\,  \m_s, \,  \n_s} \, .
\ee 

\section{Equations of motion and current exchanges} \label{sec4}
The equations of motion in the presence of a source, for the effective bosonic theory defined in \eqref{geomL}, 
read
\be \label{eombose}
\fr{1}{\Box^{s - 1}} \, \prd^{\, s} \, \cR^{\, (s)} \, (\vf)\, = \, \cJ \, .
\ee 
As anticipated in the Introduction, it is interesting to observe that fermionic, non-local equations of motion, easily computed 
from \eqref{geomL},  can be formally
obtained from the corresponding bosonic ones simply interpreting the field
as carrying an additional spinor index, while also acting on the resulting tensor with the operator $\fr{\not \, \pr}{\Box}$:
\be \label{eombosefermi}
\fr{1}{\Box^{s - 1}} \, \prd^{\, s} \, \cR^{\, (s)} \,(\vf)\, = \, \cJ 
\hspace{1cm} \ra \hspace{1cm}
\fr{\dsll}{\Box^{s }} \, \prd^{\, s} \, \cR^{\, (s)} \, (\psi)\,= \, \th \, .
\ee 

 It should be stressed that such a strong similarity between bosonic and fermionic equations
is not met in the comparison of theories describing \emph{irreducible} spins $s$ and $s + \12$.
Indeed, with the exception of the pair $(s_{B}, s_{F}) = (0, \12)$, the formal relation between bosonic and
fermionic wave operators \emph{does not} amount to the simple operations we indicate in \eqref{eombosefermi}, 
neither in the constrained (Fang-) Fronsdal case \cite{fronsdal} (or in their minimal local counterparts
\cite{fs3}), nor in the non-local formulations of \cite{fs1, fs2, fms1}.
For instance, as already noticed in \cite{fs1}, the formal relation between the Fang-Fronsdal tensor
$\cS \, = \, i\, (\dsll \, \psi - \pr\,  \psisl)$, defining the equations of motion for constrained 
fermions, and the Fronsdal one \eqref{fron},  is
\be
{\cal S} \, - \, \frac{1}{2} \, \frac{\pr}{\Box}\, {\not{\!\pr}} \, 
\ssl \ = \ i \ \frac{\not{\!\pr}}{\Box} \, {\cal F} \, (\psi) \ ,
\ee
while similar complications are met when comparing the geometric, non-local,
equations for irreducible bosons and fermions of \cite{fs1, fs2, fms1}. 

 Restricting the attention for simplicity to the bosonic case,  we would now like to show that 
the equation we found  bears the same physical content as \eqref{triplet0}. Indeed, \eqref{eombose} in the absence  
of sources can be written in the form
\be \label{eombosefield2}
\Box \, \vf \, + \,  \pr \, \{\sum_{m = 1}^{s} \, \fr{(-1)^m}{m} \, 
\fr{\pr^{\, m - 1}}{\Box^{\, m - 1}} \, \prd^m \vf \} = \, 0,
\ee 
where in particular the gauge transformation of the second term is
\be \label{gauge}
\d \, \{\sum_{m = 1}^{s} \, \fr{(-1)^m}{m} \, 
\fr{\pr^{\, m - 1}}{\Box^{\, m - 1}} \, \prd^m \vf \} \, = \, - \, \Box \, \L \, .
\ee
Thus, solving for the parameter $\L$,  it is possible to choose a gauge where 
\be \label{semifierz}
\begin{split}
& \sum_{m = 1}^{s} \, \fr{(-1)^m}{m} \, 
\fr{\pr^{\, m - 1}}{\Box^{\, m - 1}} \, \prd^m \vf  \, = \, 0 \, ,
\end{split}
\ee
and then notice that this condition  implies
$
\prd \vf \, = \, 0 \, \ft{Equivalently, we could directly fix a  gauge where $\prd \vf = 0$ solving
for $\Box \, \L + \pr \, \prd \L + \prd \hat{\vf} = 0$ by means of \eqref{inverseAapp}, for $n = 1$ and $b = 1$.}
$.
In this fashion, \eqref{eombosefield2} is reduced to the system
\be \label{quasifierz}
\begin{split}
&\Box \, \vf \, = \, 0 \, , \\
&\prd \vf \, = \, 0 \, , 
\end{split}
\ee
describing  indeed the same content as \eqref{triplet0} \ft{Adding to \eqref{quasifierz} the condition of vanishing trace, $\vf^{\, \pe} = 0$, would complete the Fierz system \cite{fierz} for irreducible propagation of massless spin $s$. In the absence of the trace condition the theory is still unitary, but all lower-spin components in $\vf$ propagate.}. The 
reliability of these manipulations  is confirmed by the analysis of the current exchange between  conserved sources\ft{See \cite{fms1, fms2} for the irreducible case, and \cite{ftcurrents} for the case of triplets. },
that in the case under scrutiny takes the extremely simple form
\be \label{currexch}
\cJ \cdot \{\Box \, \vf \, + \,  \sum_{m = 1}^{s} \, (-1)^m \, 
\fr{\pr^{\, m}}{\Box^{\, m - 1}} \, \prd^m \vf \}\, = \, 
\cJ \cdot \Box \, \vf\, = \,  \cJ \cdot \cJ \, , 
\ee
thus reproducing, as expected, the corresponding result that one obtains from the triplet. In that context in fact, eliminating $C$, the equation for $\vf$ when coupled to a source reads\ft{In terms of the diagonal basis of Fronsdal tensors, eq. \eqref{tripletJ} implies that the various propagating fields are coupled to traces of a single source, rather than to independent currents.} 
\be \label{tripletJ}
\Box \, \vf \, - \, \pr \, \prd \vf \, + \, 2 \, \pr^{\, 2} \, D \, = \, \cJ \, ,
\ee
while the equation for $D$ only provides a consistency condition for the conservation of $\cJ$. Contraction with a divergenceless source leads then to \eqref{currexch}. 
 
 At any rate, as we argued in \cite{dariocrete} for the irreducible case,
our opinion is that the very procedure we followed to derive the equations here presented provides in a sense a proof of their physical content, allowing to trace back in an unambiguous fashion all non-localities to the presence of auxiliary fields in the corresponding  local Lagrangians.

%%%%%%%%%%%%%%%%%%%%%%%%%%%%%%%%%%%%%%%%%%%%%%%%%%%%%%%%%%%%%%%%%%%%%

\section*{Acknowledgments}

%%%%%%%%%%%%%%%%%%%%%%%%%%%%%%%%%%%%%%%%%%%%%%%%%%%%%%%%%%%%%%

I am grateful to A. Campoleoni, J. Mourad and  A. Sagnotti for useful discussions. 
 For their nice hospitality extended to me while this work was in progress, I would also like to thank the Scuola Normale Superiore of Pisa 
 and the Institute of Physics  of the ASCR in Prague. The present research 
was supported  by APC-Paris VII and  by the CNRS through the P2I program, and also in part by  the MIUR-PRIN contract 2007-5ATT78 and
by the EURYI grant GACR  EYI/07/E010 from EUROHORC and ESF. \vskip 12pt

%%%%%%%%%%%%%%%%%%%%%%%%%%%%%%%%%%%%%%%%%%%%%%%%%%%%%%%%%%%%%%%%%%%%%

%%%%%%%%%%%%%%%%%%%%%%%%%%%%%%%%%%%%%%%%%%%%%%%%%%%%%%%%%%%%%%%%%%%%%

\section*{Appendix} \label{app}

%%%%%%%%%%%%%%%%%%%%%%%%%%%%%%%%%%%%%%%%%%%%%%%%%%%%%%%%%%%%%%%%%%%%%

 We would like to prove formula \eqref{inverseA}, according to which the inverse of the operator
\be
A  =  \Box^{\, n} \, (1 \, + \, b \, \fr{\pr}{\Box} \, \pr\, \cdot) 
\ee
on $T_{\, (k)}$, the space  of tensors of rank $k$, is given by the expression
\be \label{inverseAapp}
A^{\, -1}_{\, (k)} \, = \, \fr{1}{\Box^{\, n}} \, \{1\, + \, \sum_{m = 1}^k \, (-1)^m \, 
\fr{m! \, b^{\, m}}{\prod_{l = 1}^m \, (1\, + \, l \, b)} \, \fr{\pr^{\, m}}{\Box^{\, m}} \, 
\pr\, \cdot^m\}\, .
\ee
The basic observation is that on  $T_{\, (k)}$ the operator
$
P_k \, \equiv \, \fr{\pr^{\, k}}{\Box^{\, k}} \, \pr\, \cdot^k \, 
$
is idempotent:
$
P_k^{\, 2} \, = \, P_k \, ;
$
this implies that the inverse of $A$  shall contain only the operators
$P_i$ with $i \leq k$,  with coefficients that we would like to determine.
For instance on $T_{\, (1)}$ we would have
\be
A ^{\, -1}_{\, (1)} \, = \, \Box^{\, - n} \, (1 \, + \, \m \, \fr{\pr}{\Box} \, \pr\, \cdot) \, , 
\ee
from which, requiring that $A ^{\, -1}_{\, (1)} \, A \, = \, \mathbb{I}$, we obtain  $\m \, = \, - \, \fr{b}{1 + b}$.
Similarly, on $T_{\, (2)}$ we try
\be
A ^{\, -1}_{\, (2)} = \, \Box^{\, -n} \, (1 \, + \, \m_1 \, \fr{\pr}{\Box} \, \prd \, + \, 
\m_2 \, \fr{\pr^{\, 2}}{\Box^{\, 2}} \, \prd \pr\, \cdot) \, , 
\ee
and imposing  
$A ^{\, -1}_{\, (2)} \, A =  \mathbb{I}$ we find
$\m_1 =  - \, \fr{b}{(1 \, + \, b)}$ , 
$\m_2 =  + \, \fr{2\, b^{\, 2}}{(1 \, + \, b) \, (1 \, + \,2\,  b)}$.
According to the suggested pattern, we thus conjecture the general solution  to be given by \eqref{inverseAapp},
which can be proven by induction. Indeed, imposing  $A \, A^{\, -1}_{\, (k + 1)} \, = \, \mathbb{I}$ on $T_{\, (k + 1)}$, we find
the same system  as for the k-th step, whose solution is assumed to give $A^{\, -1}_{\, (k)}$, together with
terms involving $P_{k + 1}$, whose coefficients are
\be
(-1)^{\, k+1} \, \fr{ b^{\, k + 1}}{\prod_{l = 1}^{k + 1} \, (1\, + \, l \, b)}\, 
\{ (k + 1)! \, - \, (k+1) ! \, (1 + (k + 1) \, b) \, + \, b \, (k + 1)!\, (k + 1)\}\, ,
\ee
and can be easily verified to sum up to zero.

\end{document}